\documentclass[nostrict,journal=jpccck,manuscript=article,layout=twocolumn]{achemso}

\makeatletter
\let\achemso@titlefont\titlefont   
\let\titlefont\relax               
\usepackage[fontsize=9pt]{scrextend}
\let\titlefont\achemso@titlefont   
\makeatother

\usepackage[font={small}]{caption}

\usepackage{amssymb} 
\usepackage{graphicx} 
\usepackage{xcolor} 
\usepackage{float}
\usepackage{mhchem} 
\usepackage{siunitx}
\usepackage{lipsum} 
\usepackage{epsfig}
\usepackage{ragged2e}
\usepackage{hyperref}
\usepackage{bm}
\usepackage{amsmath,amsfonts,amsthm,bm}
 
\DeclareUnicodeCharacter{2212}{$-$}

\graphicspath{{./figures/}}

\author{Marco Fornari}
\affiliation{Department of Physics, Central Michigan University, Mount Pleasant MI, USA}
\author{Virginia Carnevali}
\affiliation{Physics Department, University of Trieste, Trieste, Italy}
\email{virginia.carnevali@units.it}

\title[An \textsf{achemso} demo]
  {Microscopic structure at the morphotropic phase boundary in lead-barium meta-niobate}

\abbreviations{IR,NMR,UV}

\begin{document}
\begin{abstract}
Lead–barium niobate Pb$_{1-x}$Ba$_x$Nb$_2$O$_6$ is a tetragonal tungsten‑bronze ferroelectric that exhibits enhanced functional properties near a morphotropic phase boundary separating orthorhombic and tetragonal phases. Using first‑principles electronic‑structure calculations combined with ab initio molecular dynamics, we investigate how Ba substitution modifies cation-site occupancy, local distortions, and polarization behavior across the solid solution. By explicitly accounting for vacancy ordering and the inequivalent $A$-site cages intrinsic to the tungsten‑bronze framework, we identify the structural origin of the orthorhombic‑to‑tetragonal transition. Increasing Ba content progressively weakens the cooperative in‑plane off‑centering of Pb ions in the pentagonal channels and promotes out‑of‑plane displacements of Nb and $A$‑site cations, leading to a rotation of the spontaneous polarization from [110] to [001]. The transition occurs near $x \approx 0.6$, in quantitative agreement with experimental phase diagrams. 
These results provide a microscopic interpretation of the morphotropic phase boundary in Pb$_{1-x}$Ba$_x$Nb$_2$O$_6$ and establish a clear structure–polarization relationship for tungsten‑bronze ferroelectrics.
\end{abstract}

\section{Introduction}

Tungsten bronze (TB), with the general formula $A^{XII}A^{XV}A^{IX}B^{(1)}_8B^{(2)}_2X_{30}$, constitute a broad class of perovskite-like materials built from a three-dimensional network of corner-sharing $BX_6$ octahedra.\cite{lines_principles_2001} This framework generates pentagonal ($A^{XV}$), square ($A^{XII}$), and trigonal ($A^{IX}$) channels running along the crystallographic $c$-axis, enabling extensive chemical flexibility (Fig.~\ref{PN}). Owing to this versatility, TBs have attracted considerable interest as plasmonic materials,\cite{tegg_review_2023} superconductors,\cite{BIERSTEDT196625} and catalysts.\cite{KASPERA201776}

Among TBs, tetragonal tungsten bronzes (TTBs) form a distinctive class of ferroelectrics characterized by a non-pseudocubic symmetry and a reduced polarization orientation space, which can be effectively one- or two-dimensional.\cite{Lines} This feature underpins unconventional polarization rotation mechanisms and enhanced electromechanical responses. In particular, niobate-based TTBs represent a rich family of ferroelectric compounds, including (Ba$_{1-x}$Sr$_x$)$_4$Na$_2$Nb$_{10}$O$_{30}$ (SBN),\cite{https://doi.org/10.1111/j.1151-2916.1969.tb09183.x} Ba$_4$Na$_2$Nb$_{10}$O$_{30}$ (BNN),\cite{londal_structural_2024}  and related mixed-cation systems,\cite{Jiang1998SingleCG,Randall} whose fundamental properties arise from the interplay of high Curie temperatures,\cite{khesro_temperature_2016} structural complexity and the possibility of site-selective substitution within the $A$-site cages,\cite{Sahini} and the occurrence of morphotropic phase boundaries (MPBs).\cite{MPB}

One important motivation for studying TTBs lies in their potential as tunable functional materials for optoelectronic applications, including their possible use as active layers in photovoltaic devices. The TTB framework, with its multiple inequivalent $A$-site cages and inherent structural flexibility, offers a unique platform for band gap engineering through selective cation substitution and compositional control.\cite{D4NR04655B} In analogy with oxide and hybrid perovskites, where electronic states near the band edges are primarily governed by the $B$-$X$ network while $A$-site cations modulate structural stability and indirectly influence electronic properties,\cite{C6TA04949D} similar design principles can be extended to TTB systems. Once a suitable structural configuration is achieved, the band gap can be systematically tuned by varying chemical composition at both $A$- and $B$-sites. Moreover, the presence of distinct cages ($A^{XII}$, $A^{XV}$, and $A^{IX}$) opens the possibility of incorporating molecular or organic species into the framework, thereby introducing additional degrees of freedom to tailor the electronic structure, dielectric response, and light–matter interactions.

Within TTBs, lead–barium niobate, Pb$_{1-x}$Ba$_x$Nb$_{10}$O$_{30}$ (PBN) that prototypicales between PbNb$_2$O$_6$ (PN) and BaNb$_2$O$_6$ (BN), exhibiting a MPB at $x\approx0.63$.\cite{GOODMAN} Compositions near the MPB are known to display enhanced and weakly temperature-dependent dielectric, piezoelectric, and pyroelectric responses, making them particularly attractive for functional applications. \cite{Shrout,Shrout1,Lane}  Upon cooling, niobate TTBs undergo a transition from a paraelectric tetragonal phase to either a non-centrosymmetric tetragonal (TET) or an orthorhombic (ORT) ferroelectric phase.\cite{https://doi.org/10.1111/j.1151-2916.1989.tb06102.x} The ferroelectric ground state arises from a cooperative Jahn–Teller-like distortion involving off-centering of both $A$- and $B$-site cations and octahedral tilts of the form  $a^0a^0c^{\pm}$.\cite{olsen_role_2017}  In PBN, the MPB separates an orthorhombic phase with polarization along [110] from a tetragonal phase polarized along [001], enabling polarization rotation through modest changes in composition or temperature.\cite{Cross,Guo}

Despite extensive experimental and theoretical investigations,\cite{Guo1,Hyde} several key aspects of PBN remain unresolved. Previous studies have largely focused on macroscopic structure–property relationships and average crystallographic descriptions,\cite{roshchupkin_study_2009} often relying on simplified models that do not explicitly account for the intrinsic complexity of the TTB lattice,\cite{Francombe,Francombe2} including the coexistence of multiple inequivalent $A$-site cages, partial occupancies, and vacancy ordering.\cite{Olsen} As a result, the microscopic origin of the morphotropic phase boundary, as well as the role of local cation distribution and lattice distortions in governing polarization rotation, is still not well understood.

In this work, we address these limitations by adopting a fully atomistic, first-principles approach that explicitly captures (i) the inequivalent $A$-site environments, (ii) realistic Pb/Ba distributions across the different cages, and (iii) vacancy ordering within the TTB framework. By removing symmetry constraints typically imposed in reduced models and systematically exploring representative configurations across composition, we provide a microscopic description of how local chemical environments control structural distortions and polarization behavior. This approach enables us to (i) identify the atomistic mechanism driving the orthorhombic-to-tetragonal transition, (ii) establish a direct link between site-specific chemistry and polarization rotation, and (iii) clarify the extent to which electronic properties near the band edges are affected by cation substitution. More broadly, our results provide general design principles for tuning functional properties in non-perovskite ferroelectric frameworks and highlight the potential of TTBs as chemically flexible platforms for advanced electro-optic and photovoltaic applications.

\section{Results and discussion}
The section is structured as follows. The first paragraph discusses the rationale behind the choice of the simulation cell that most accurately reproduces the PBN atomic structure. In particular, it addresses the doubling of the cell along the $c$-direction, the choice of the in-plane cell dimensions, the arrangement of the unoccupied $A^{XII}$ sites within the cell, and the resulting model zoo generated by Pb–Ba substitutions at the $A^{XV}$ sites. The second paragraph examines the energetic and electronic properties as a function of Ba concentration, including, where relevant, comparisons with analogous perovskite systems. Finally, the third paragraph focuses on the ORT–TET phase transition, with particular attention to the MPB.

\begin{figure}[h!] 
  \centering
  \includegraphics[width=0.48\textwidth]{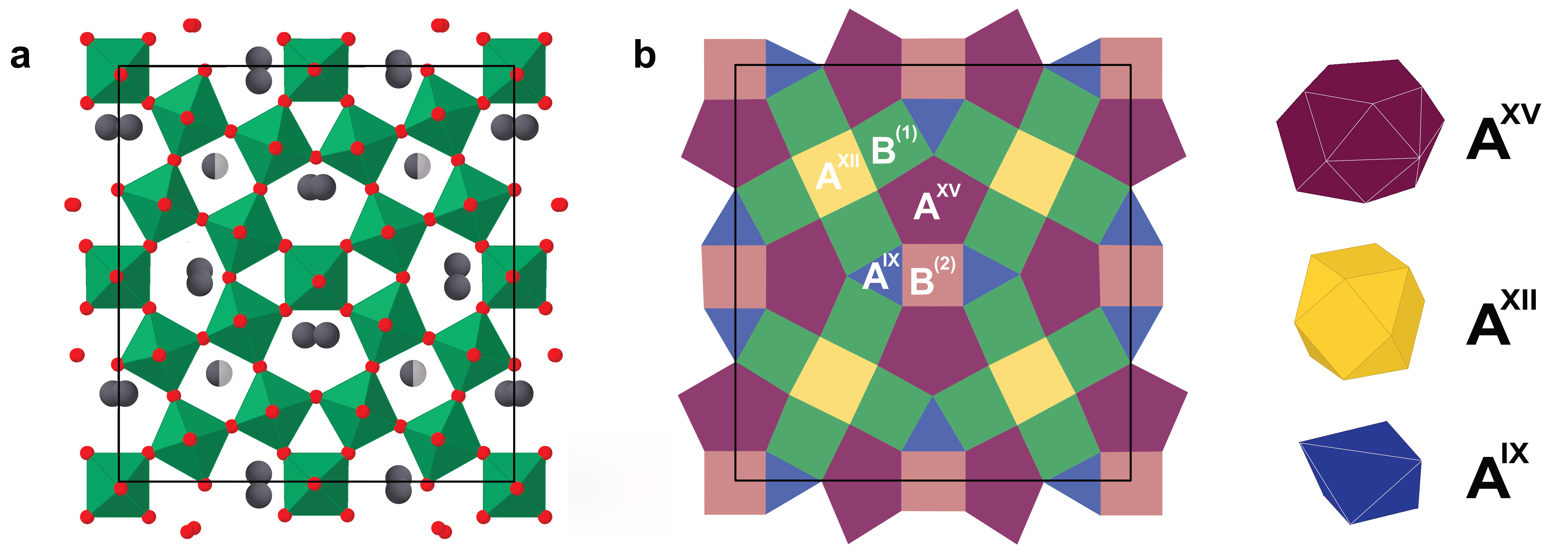}
  \caption{(a) Top view of the PN TET structure. Pb, O, and Nb are represented by gray balls, red balls, and green octahedra, respectively. The computational cell is doubled along the $c$-direction. $A^{XII}$ sites are occupied by Pb. This arrangement is implemented to ensure the preservation of the system's electro-neutrality. Here we show the most stable configuration, wherein the $A^{XII}$ are arranged in a checkerboard pattern, with sites occupied in one layer being unoccupied in the other. (b) Same cell with the three inequivalent $A$-sites ($A^{XV}$, $A^{XII}$, and $A^{IX}$) and the two $B$-sites ($B^{(1)}$ and $B^{(2)}$) highlighted. $A^{XV}$, $A^{XII}$, and $A^{IX}$ three dimensional cages are shown.}
  \label{PN}
\end{figure}

\paragraph{Atomic structure.}
Owing to their ionic radii (1.61 Å for Ba$^{2+}$ and 1.49 Å for Pb$^{2+}$),\cite{Shannon} Ba preferentially occupies the larger $A^{XV}$ sites, whereas Pb can populate both $A^{XV}$ and $A^{XII}$ channels. Starting from the experimental orthorhombic structure\cite{Labbe}, we reconstruct the prototype tetragonal phase for PN (Fig.S1). The computational cell contains 180 atoms and, due to Pb vacancies in four $A^{XII}$ sites to preserve electro-neutrality, has P4/m symmetry (Fig.\ref{PN}).
The presence of many instabilities in the out-plane direction suggests that the doubling along the $c$-direction may be energetically favored. Our analysis indicates that the doubling is energetically preferred, enabling Jahn-Teller distortion for the octahedra in the $c$-direction as it has been already shown for other TTB compositions\cite{gardner_ferroelectricity} and as it should be for the perovskite-like materials\cite{carnevali_nanoscale_2025}. Rigid octahedra rotations around the $c$-axis are constrained by structural limitations.

From a pure geometrical point of view our simulation cell is a conventional cell where a single TTB layer contains four $A^{XII}$ and eight $A^{XV}$ sites -- 180 atoms, hereafter referred to as cell-1 -- while the primitive cell should have two $A^{XII}$ and four $A^{XV}$ sites -- 90 atoms, hereafter referred to as cell-2 (Fig.S1). This hypothesis was initially postulated; however, it was subsequently invalidated due to the inability to accurately capture the ORT-TET phase transition when Pb is substituted by Ba.
Cell-2 imposes the base-center orthorhombic symmetry on cell-1, treating as independent only two $A^{XII}$ and four $A^{XV}$ among the four and eight sites present in cell-1, being a suitable subject for the study of PN; the prototype tetragonal structure relaxes to orthorhombic, with an energy gain of 60 meV/atom, exhibiting a strong correlation with experimental findings.\cite{GOODMAN}
However, complications emerge when considering the PBN composition.
The Pb-Ba substitution disrupts the PN base-center orthorhombic symmetry, potentially resulting in the non-equivalence of some of the $A^{XV}$ occupied sites. The position of Pb inside the A$^{XII}$ cage will differ depending on the occupation of the $A^{XV}$ sites (Pb rather than Ba) in the surrounding cages. This mechanism is fundamental to a proper description of the ORT-TET phase transition in PBN. Therefore, it is imperative to release all symmetry constraints on $A^{XII}$ and $A^{XV}$ and cell-1 is the optimal choice for this purpose. 
With cell-2, the relaxed structure never ends up having tetragonal symmetry when all the $A^{XV}$ cages are filled with Ba. This underlies that the ORT-TET is a complex phase transition where all the four $A^{XII}$ and eight $A^{XV}$ sites have to be involved without additional symmetry constrains. Not all the TTB compounds need a cell-2-like model to be properly described. As a title of example, the ferroelectric transition in SBN is fairly described by a cell-1-like model.\cite{Olsen} The need of the cell-1 for PBN may be also attributed to the complexity induced by including Pb cations in the structure.

The number of available PBN structures increases drastically depending on which $A^{XII}$ cage is filled, the Ba concentration, and the $A^{XV}$ occupation. As previously indicated, the PBN stoichiometry suggests the occupation of two out of four $A^{XII}$ cages. Preliminary calculations indicate that the system favors a configuration that maximizes the distance in both the in-plane and $c$-direction between $A^{XII}$ vacancies. The configurations that have been selected feature in-plane vacancies organized in a checkerboard pattern. A screw axis extending along the $c$-direction rotates the vacancy pattern of the second layer by 90° with respect to the one in the first layer (Fig.S2). The relaxed structures with this vacancy arrangement exhibit, on average, 50 meV/atom more stability than all the other possible structures with different vacancy arrangements (Tab.S1). Henceforth, all models under consideration will maintain this fix configuration for the $A^{XII}$ sites.

The PN structure ($x = 0$) displays a single configuration, a consequence of the complete saturation of all $A^{XV}$ cages by Pb. This also applies to $x = 0.8$ because all the A$^{XV}$ cages contain Ba. A far more complex scenario arises with the coexistence of $A^{XV}$ filled by Pb and Ba ($x = 0.2, 0.4, 0.6$). The only constraint imposed on the system is that the Ba concentration in each layer must be equivalent. Consequently, structures exhibiting Ba accumulations in one of the two layers are not permitted. The single layer has eight available $A^{XV}$ cages, in which a given number of Ba, $n_x < 8$, according to the concentration $x$, can be accommodated. Upon consideration of the internal symmetry resulting from the vacancy checkerboard pattern, and subsequent rejection of equivalent configurations resulting from rigid translations of a single lattice vector along the $c$-direction, the total number of configuration at a given Ba concentration $x$ is:
\begin{equation}
  N_x = \left(\frac{1}{2}\frac{8!}{n_x!(8-n_x)!}\right)^2,
  \label{EqMod}
\end{equation}
where $n_x$ is the number of Ba in the cell.
Figure S3 summarizes the configurations at different Ba concentrations.
Similar to the 3D perovskites,\cite{Pero} TTB structural stability should satisfy some empirically derived tolerance factors given by the ionic radii of the cations and anions located in the $A$ ($r_A$), $B$ ($r_B$), and $X$ ($r_X$) sites. The square cage stability is checked by the Goldschmidt tolerance factor $t_{XII}$,\cite{Gold} while for the pentagonal cage a new tolerance factor $t_{XV}$ has been derived taking into account the geometry of the $A^{XV}$ cage. Hence, the TTB tolerance factors are:
\begin{equation}
  t_{XII} = \frac{r_A + r_X}{\sqrt{2}(r_B + r_X)}, \qquad
  t_{XV} = \frac{r_A + r_X}{r_B + r_X} \sin{\pi/5}.
  \label{TF}
\end{equation}
It has been shown that the stability of the perovskite structure may be expected for $0.71 < t_{XII} < 1.00$.\cite{Pero,toll} According to these criteria, PBN are stable structures, with $t_{XII} = 0.90(0.95)$ and $t_{XV} = 0.77(0.79)$ for Pb-rich(Ba-rich).

\paragraph{Electronic properties.}
The TTB structures were optimized and the formation energies for the two configurations at $x = 0.0, 0.8$ were computed. In addition, the formation energies for a subset of configurations at $x = 0.2, 0.4, 0.6$ were also computed. The formation energy is computed with respect to the typical building blocks used for the PBN synthesis -- BaO, PbO, and Nb$_2$O$_5$ -- according to the following equation:
\begin{equation}
  E_f = [E - n_x E_{\text{BaO}} - (20 - n_x)E_{\text{PbO}}-20 E_{\text{Nb}_{2}\text{O}_{5}}] / n_{atm},
\end{equation}
where $E$ is the total energy of the optimized TTB structure; $n_x$ is the number of Ba atoms in the cell at given concentration $x$; $E_{\text{BaO}}$, $E_{\text{PbO}}$, and $E_{\text{Nb}_{2}\text{O}_{5}}$ are the total energies for BaO, PbO, and Nb$_2$O$_5$ relaxed bulk structures, respectively; $n_{atm}$ is the number of the atoms in the cell. At a given Ba concentration, the formation energy surface is relatively level (Tab. S1, with a maximum variation of 1 meV per atom). This finding suggests a high probability of coexisting configurations, thereby validating the decision to study a limited configuration subset for each Ba concentration without compromising the generality in PBN characterization.

\begin{figure}
  \centering
  \includegraphics[scale=0.3]{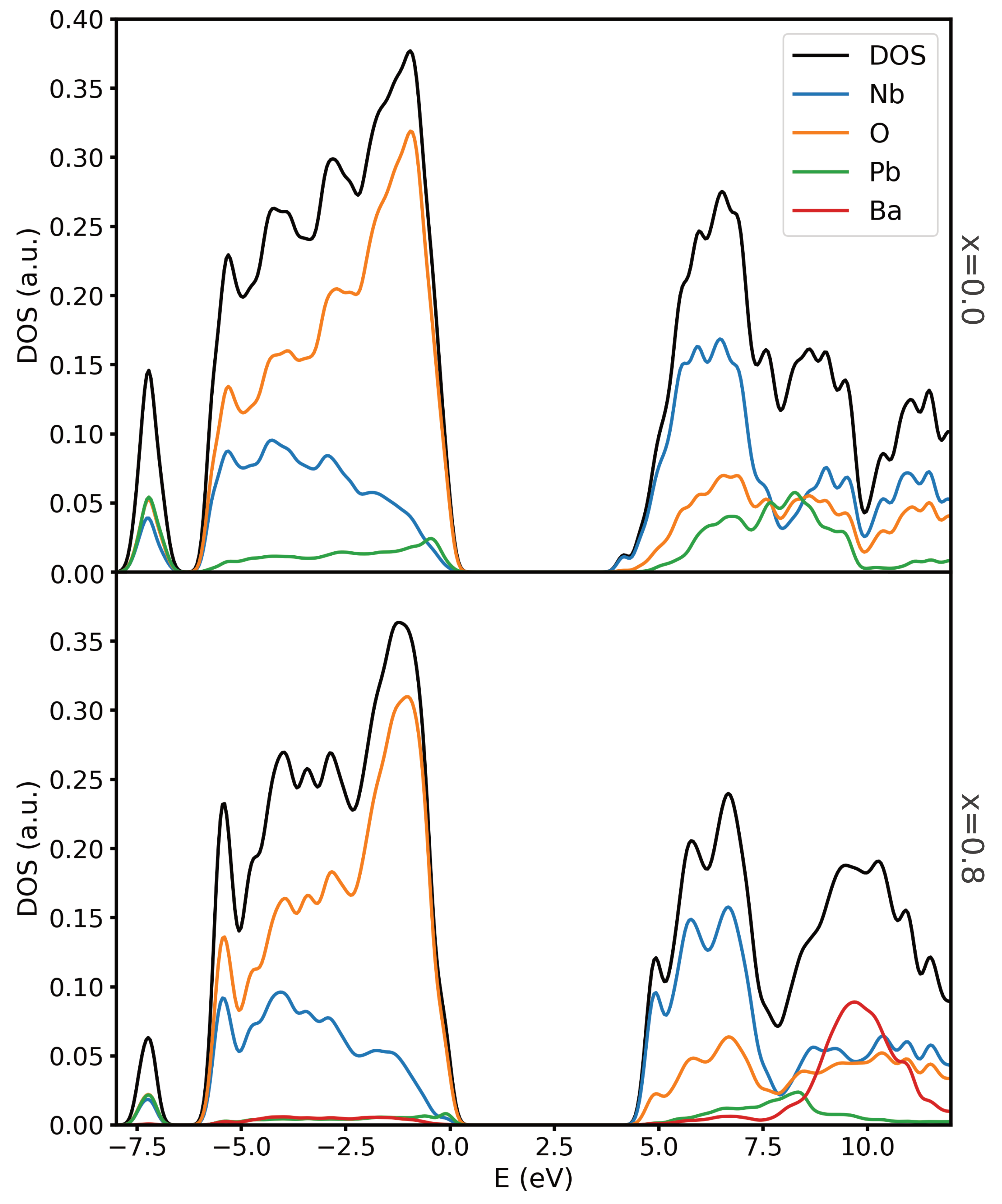}
  \caption{Density of states (DOS) and projected density of states (PDOS) on the different atomic species for Pb-rich (top panel, $x=0.0$) and Ba-rich (bottom panel, $x=0.8$) structures. The Fermi energy is set to 0 eV. The simulations were performed at PBE0 level of theory.}
  \label{DOS}
\end{figure}

Figure \ref{DOS} shows the projected density of states (PDOS) of $x = 0.0, 0.8$ structures were perovskite-like features can be noted.\cite{D0RA06028C} The top of the valence band (TVB) and bottom of conduction band (BCB) displays hybridization between O and Nb, a typical interaction for elements that are organized in perovskite-like octahedral networks.\cite{C6TA04949D}  A trace amount of Pb is present in the TVB and BCB, whereas Ba exhibits a negligible presence. This observation can be attributed to the distinct filling patterns of the $A^{XII}$ and $A^{XV}$ cages by Pb and Ba, respectively. Specifically, Pb has the capacity to fill both cages, while Ba is capable of occupying solely the $A^{XV}$ cage. Due to the smaller size of the $A^{XII}$ cage, the Pb in this cage will be spatially closest to the O atoms with respect to the $A^{XV}$ cage ($d_{Pb-O}^{XII}=2.45$ \AA{}, $d_{Pb-O}^{XV}=2.75$ \AA{} Fig.S4). This spatial proximity is reflected in a slight hybridization between Pb and O. Furthermore, Ba exhibits a less off-centering motion relative to Pb, which can be attributed to the larger ionic radius of Ba ($d_{Ba-O}^{XV}=2.95$ \AA{}, Fig.S5).  The Pb-Ba substitutions have a negligible effect on the electronic states that are proximate to the band gap. This finding is consistent with the established characteristics of perovskite compounds, wherein the $A$ site serves to stabilize the structure while maintaining electronic states in proximity to the band gap.\cite{PeroRev,Halide}

\begin{figure}
  \centering
  \includegraphics[scale=0.28]{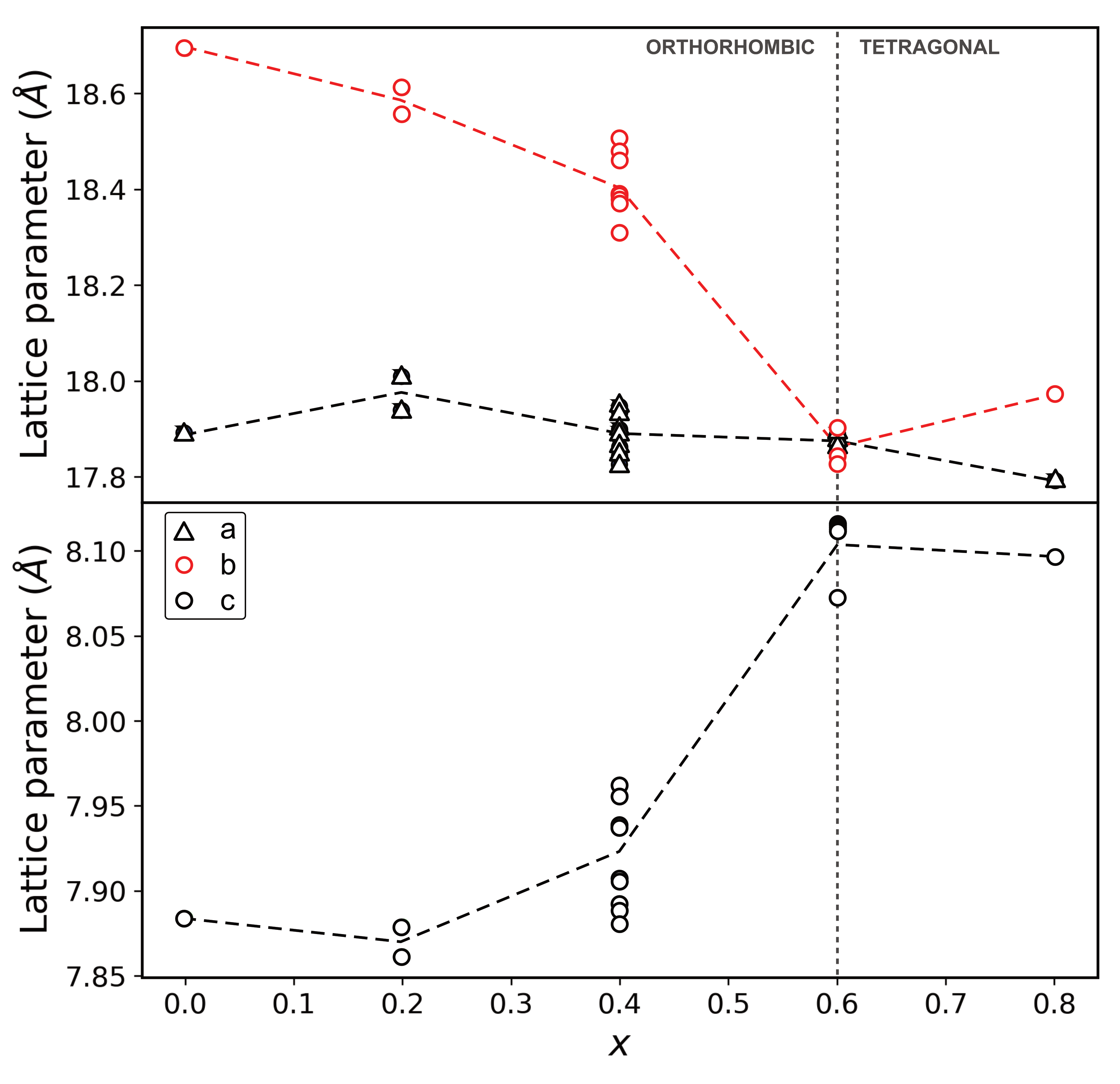}
  \caption{Lattice parameters obtained after relaxing the atomic structures at each Ba concentration $x$. The plot of the a, b parameters clearly shows a transition between ORT ($a \not= b$) and TET ($a = b$) at $x \approx 0.6$; the increasing of the parameter $c$ with $x$ agrees with the rotation of the polarization from the [110] direction (ORT) to the [001] direction (TET).}
  \label{LP}
\end{figure}

\paragraph{MPB.}
Figure \ref{LP} illustrates the ORT-TET phase transition for the PBN, as a function of the Ba concentration $x$ within the structure. The difference of the in-plane lattice parameters $a$ and $b$ tends to zero with the increasing of Ba concentration, leading the symmetry of the system from orthorhombic to tetragonal with a MPB at $x \approx 0.6$, as expected.\cite{GOODMAN} The slight discrepancy between $a$ and $b$ at $x = 0.8$ can be attributed to the proximity of PBN to a region where the system exhibits hexagonal symmetry ($x = 1$). The Nb and Pb atoms in the PbNb$_2$O$_6$ compound exhibit only in-plane displacements (Fig.S6, Tab.S2). The observed increase in the lattice parameter $c$ is indicative of the displacement of the Nb, Pb, and Ba atoms along the $c$ direction. This hypothesis is corroborated by a statistical analysis of NVE trajectories at 300 K, which reveals an off-centering along the $c$-direction of the Nb and Pb atoms when Ba is also included in the structure. It is important to note that Pb atoms show also an out-off-plane displacement component only in presence of Ba (Fig.S4-S5, Tab.S2). This phenomenon can be attributed to a symmetry-breaking effect caused by Ba, which lowers the cooperative in-plane interaction between Pb atoms. This, in turn, decreases the Pb-O interaction, thereby allowing the out-of-plane motion of Nb.
The displacement of Pb and Ba along the $c$-direction within the $A^{XV}$ and $A^{XII}$ cages is comparable in magnitude and opposite in direction to the displacement of Nb within the NbO$_6$ octahedra (Fig. S7). However, due to the higher oxidation state of Nb$^{5+}$ compared to that of Pb$^{2-}$ and Ba$^{2-}$, this results in a polarization along the $c$-direction. It can be posited that the introduction of Ba in the Pb$_{1-x}$Ba$_x$Nb$_2$O$_6$ solid solution results in spontaneous polarization switching from in-plane to out-of-plane, with a MPB at $x \approx 0.6$, where the tetragonal symmetry is reduced to orthorhombic (Fig.\ref{LP}). A statistical analysis of NVE trajectories at 300 K yielded an in-plane polarization of 15.5 $\pm$ 0.02 $\mu$C/cm$^2$ for PbNb$_2$O$_6$ and out-of-plane polarization of 19.9 $\pm$ 0.02 $\mu$C/cm$^2$ for Pb$_{0.23}$Ba$_{0.67}$Nb$_2$O$_6$.
As demonstrated previously for PbTiO$_3$,\cite{PT,PT1,MF} the aforementioned characteristics support the possibility of cooperative motion between $A$ and $B$ sites in the ORT-TET phase transition.

\section*{Conclusions}
We have elucidated the microscopic mechanism governing polarization switching across the MPB in Pb$_{1-x}$Ba$_x$Nb$_2$O$_6$ by combining first‑principles calculations with AIMD and an explicit treatment of cation-site occupancy and vacancy ordering in the TTB lattice. In Pb‑rich compositions, the orthorhombic phase is stabilized by a cooperative in‑plane off‑centering of Pb ions in the pentagonal channels together with correlated Nb displacements within the NbO$_6$ octahedra, giving rise to a spontaneous polarization along [110]. Progressive substitution of Pb by Ba disrupts this cooperative in‑plane distortion, lowers the Pb–O interaction, and enables significant out‑of‑plane displacements of both Nb and $A$‑site cations. This crossover drives a continuous rotation of the polarization from [110] to [001] and stabilizes the tetragonal phase near $x \approx 0.6$, in agreement with the experimentally observed morphotropic phase boundary.

The electronic structure calculations shows that Ba substitution has a negligible impact on the band edges, confirming that the morphotropic behavior is controlled by local structural rearrangements rather than by changes in electronic hybridization. Our results highlight the central role of site‑specific chemistry and cooperative $A$–$B$‑site displacements in controlling phase stability and polarization orientation in TB ferroelectrics. Beyond Pb$_{1-x}$Ba$_x$Nb$_2$O$_6$, this work provides general microscopic guidelines for designing and tuning MPB in non‑perovskite ferroelectric frameworks to optimize functional responses.

\section*{Methods}
\paragraph{Electronic properties at 0 K.}
Ab-initio electronic calculations at 0 K were performed using Quantum ESPRESSO\cite{giannozzi_quantum_2009} as implemented in AFLOWP$\pi$\cite{supka_aflow_2017}. We used DOJO norm-conserving PBE pseudopotentials \cite{van_setten_pseudodojo_2018} with well-converged basis sets corresponding to an energy cutoff of 150 Ry for the wave functions and 600 Ry for the charge density. A subsequent evaluation of the spin-orbit coupling in the calculations yielded an inconsequential contribution to the electronic properties of the systems. This observation is attributable to the absence of participation by heavy atoms such as Pb and Ba in chemical bonds. To integrate over the Brillouin zone, for TTB structures, PbO and BaO, and Nb$_2$O$_5$, we employed $3\times3\times6$, $12\times12\times12$, and $6\times6\times6$ Monkhorst-Pack k-point grids\cite{monkhorst_special_1976}, respectively. DOS and PDOS were computed on single-point energy calculations at PBE0 level of theory performed on the optimized PBE structures. For each concentration of Ba, a sample of representative structures was generated in which the Ba positions in the $A^{XV}$ cages were assigned randomly.

\paragraph{Born-Oppenheimer ab-initio molecular dynamics.} 
AIMD simulations were performed with the mixed Gaussian and plane wave (GPW) method, as implemented in the CP2K package.\cite{kuhne_cp2k_2020} Valence electrons were expanded as a double-$\zeta$ Gaussian  basis set with polarization functions (DZVP)\cite{vandevondele_gaussian_2007}. The energy cutoff for the electron density expansion in the GPW method was 1200 Ry.
The systems were equilibrated at 300 K and 1 atm under isobaric ensemble (NPT) where the temperature was controlled by the Bussi thermostat \citep{bussi_canonical_2007} and the pressure by the Martyna barostat.\citep{martyna_constant_1994}  The systems were than equilibrated in the microcanonical ensemble (NVE), and statistics were gathered for the last 5 ps of the production run. All the MD simulations were run using an integration time step of 0.5 fs and a $1\times1\times2$ supercell. 

\paragraph{Macroscopic polarization.} 
The macroscopic polarization was evaluated from local atomic displacements using a point-charge model based on formal oxidation states. Statistics were calculated from a subset of snapshots sampled every 25 fs from AIMD NVE trajectories. The displacement of each atom was defined relative to either the centre of its local coordination environment or the corresponding electronic centers.
Specifically, the displacement of each Nb atom was computed as the vector difference between the instantaneous Nb position and the geometric center of the six nearest oxygen atoms forming its surrounding NbO$_6$
octahedron. The octahedral center was constructed from the wrapped coordinates of the coordinating O atoms using fractional coordinates and the minimum-image convention, ensuring a consistent definition across periodic boundaries.
For the $A$-site cations Ba and Pb, the displacements were defined relative to the centers of their associated maximally localized Wannier functions.\cite{marzari_maximally_2012,PhysRevB.61.10040} For each Ba and Pb atom, the positions of the corresponding Wannier centers were identified from the electronic structure calculation and their geometric average was used as the reference electronic center. The displacement vector was then defined as the difference between the instantaneous ionic position and the center of its associated Wannier cloud, thereby accounting for the local electronic polarization around the $A$-site cations.
For O atoms, the displacement was defined relative to the midpoint of the two nearest $B$-site cations forming the corresponding $B$–O–$B$ linkage.
The polarization vector was computed as
\begin{equation}
    \mathbf{P} = \frac{1}{\Omega}\sum_k q_k \mathbf{d}_k,
    \label{TotPol}
\end{equation}
where $\mathbf{d}_k$ is the displacement of atom $k$, $q_k$ is its formal ionic charge, and $\Omega$ is the simulation cell volume. The formal oxidation states Nb$^{5+}$, Pb$^{2+}$, Ba$^{2+}$, and O$^{2-}$ were used throughout.

\section*{Acknowledgments}
V.C. acknowledges computational resources from the Swiss National Computing Centre (CSCS). V.C and M.F. acknowledge computational resources from High-Performance Computing Center at Michigan State University. We thank Rajendra Paudel for useful discussions.

\section*{Author contributions statement}
M.F. and V.C. conceived the idea. V.C performed the theoretical simulations. All authors contributed to the discussion and writing of the paper.

\section*{Competing interests}
The authors declare no competing interests.

\section*{Supporting Information}
The online version contains supplementary material available at \\
Data and analysis scripts are available on Zenodo at 

\bibliography{ttb-biblio}

@Article{C6TA04949D,
author ="Meloni, Simone and Palermo, Giulia and Ashari-Astani, Negar and Grätzel, Michael and Rothlisberger, Ursula",
title  ="Valence and conduction band tuning in halide perovskites for solar cell applications",
journal  ="J. Mater. Chem. A",
year  ="2016",
volume  ="4",
issue  ="41",
pages  ="15997-16002",
publisher  ="The Royal Society of Chemistry",
doi  ="10.1039/C6TA04949D",
url  ="http://dx.doi.org/10.1039/C6TA04949D"}

@Article{D4NR04655B,
author ="Kendall, Owen and Melendez, Lesly V. and Guven Bicer, Merve Nur and Wilms, Michael and van Embden, Joel and Gómez, Daniel E. and Calzolari, Arrigo and Prezzi, Deborah and Della Gaspera, Enrico",
title  ="Tungsten oxide nanocrystals doped with interstitial methylammonium cations",
journal  ="Nanoscale",
year  ="2025",
volume  ="17",
issue  ="12",
pages  ="7030-7034",
publisher  ="The Royal Society of Chemistry",
doi  ="10.1039/D4NR04655B",
url  ="http://dx.doi.org/10.1039/D4NR04655B"}

@article{https://doi.org/10.1111/j.1151-2916.1989.tb06102.x,
author = {Oliver, John R. and Neurgaonkar, Ratnakar R. and Cross, L. Eric},
title = {Ferroelectric Properties of Tungsten Bronze Morphotropic Phase Bounary Systems},
journal = {Journal of the American Ceramic Society},
volume = {72},
number = {2},
pages = {202-211},
doi = {https://doi.org/10.1111/j.1151-2916.1989.tb06102.x},
url = {https://ceramics.onlinelibrary.wiley.com/doi/abs/10.1111/j.1151-2916.1989.tb06102.x},
eprint = {https://ceramics.onlinelibrary.wiley.com/doi/pdf/10.1111/j.1151-2916.1989.tb06102.x},
year = {1989}
}

@article{khesro_temperature_2016,
	title = {Temperature stable and fatigue resistant lead-free ceramics for actuators},
	volume = {109},
	issn = {0003-6951},
	url = {https://doi.org/10.1063/1.4964411},
	doi = {10.1063/1.4964411},
	number = {14},
	urldate = {2026-01-08},
	journal = {Applied Physics Letters},
	author = {Khesro, Amir and Wang, Dawei and Hussain, Fayaz and Sinclair, Derek C. and Feteira, Antonio and Reaney, Ian M.},
	month = oct,
	year = {2016},
	pages = {142907},
}

@article{Jiang1998SingleCG,
  title={Single crystal growth and ferroelectric properties of $\alpha$(Ba1−xSrx)Nb2O6:$\beta$(Na1−yKy)NbO3 solid solutions},
  author={Yi-jian Jiang and Ruyan Guo and Amar Singh Bhalla},
  journal={Journal of Applied Physics},
  year={1998},
  volume={84},
  pages={5140-5146},
  url={https://api.semanticscholar.org/CorpusID:118957161}
}

@article{londal_structural_2024,
	title = {Structural and electrical properties of tetragonal tungsten bronze {K4Bi2Nb10O30}-{Ba4Na2Nb10O30} solid solutions across two proximate phase transitions},
	volume = {44},
	issn = {0955-2219},
	url = {https://www.sciencedirect.com/science/article/pii/S0955221924005107},
	doi = {https://doi.org/10.1016/j.jeurceramsoc.2024.05.077},
	number = {13},
	journal = {Journal of the European Ceramic Society},
	author = {Løndal, Nora Statle and Zeiger, Caren Regine and Grendal, Ola Gjønnes and Einarsrud, Mari-Ann and Walker, Julian and Grande, Tor},
	year = {2024},
	pages = {7680--7688},
}

@article{https://doi.org/10.1111/j.1151-2916.1969.tb09183.x,
author = {GIESS, E. A. and SCOTT, B. A. and BURNS, G. and O'KANE, D. F. and SEGMÜLLER, A.},
title = {Alkali Strontium-Barium-Lead Niobate Systems with a Tungsten Bronze Structure: Crystallographic Properties and Curie Points},
journal = {Journal of the American Ceramic Society},
volume = {52},
number = {5},
pages = {276-281},
doi = {https://doi.org/10.1111/j.1151-2916.1969.tb09183.x},
url = {https://ceramics.onlinelibrary.wiley.com/doi/abs/10.1111/j.1151-2916.1969.tb09183.x},
eprint = {https://ceramics.onlinelibrary.wiley.com/doi/pdf/10.1111/j.1151-2916.1969.tb09183.x},
year = {1969}
}

@book{lines_principles_2001,
	address = {Oxford, New York},
	series = {Oxford {Classic} {Texts} in the {Physical} {Sciences}},
	title = {Principles and {Applications} of {Ferroelectrics} and {Related} {Materials}},
	isbn = {978-0-19-850778-9},
	publisher = {Oxford University Press},
	author = {Lines, M. E. and Glass, A. M. and Lines, M. E. and Glass, A. M.},
	year = {2001},
}

@article{PhysRevB.61.10040,
  title = {General and efficient algorithms for obtaining maximally localized Wannier functions},
  author = {Berghold, Gerd and Mundy, Christopher J. and Romero, Aldo H. and Hutter, J\"urg and Parrinello, Michele},
  journal = {Phys. Rev. B},
  volume = {61},
  issue = {15},
  pages = {10040--10048},
  numpages = {0},
  year = {2000},
  month = {Apr},
  doi = {10.1103/PhysRevB.61.10040},
  url = {https://link.aps.org/doi/10.1103/PhysRevB.61.10040}
}

@article{marzari_maximally_2012,
	title = {Maximally localized Wannier functions: Theory and applications},
	volume = {84},
	url = {https://link.aps.org/doi/10.1103/RevModPhys.84.1419},
	doi = {10.1103/RevModPhys.84.1419},
	pages = {1419--1475},
	number = {4},
	journaltitle = {Rev. Mod. Phys.},
	author = {Marzari, Nicola and Mostofi, Arash A. and Yates, Jonathan R. and Souza, Ivo and Vanderbilt, David},
	year = {2012},
}

@article{bussi_canonical_2007,
	title = {Canonical sampling through velocity rescaling},
	volume = {126},
	issn = {0021-9606},
	url = {https://aip.scitation.org/doi/10.1063/1.2408420},
	doi = {10.1063/1.2408420},
	number = {1},
	urldate = {2022-05-30},
	journal = {The Journal of Chemical Physics},
	author = {Bussi, Giovanni and Donadio, Davide and Parrinello, Michele},
	month = jan,
	year = {2007},
	note = {Publisher: American Institute of Physics},
	pages = {014101},
}

@article{martyna_constant_1994,
	title = {Constant pressure molecular dynamics algorithms},
	volume = {101},
	issn = {0021-9606},
	url = {https://doi.org/10.1063/1.467468},
	doi = {10.1063/1.467468},
	number = {5},
	urldate = {2024-08-25},
	journal = {The Journal of Chemical Physics},
	author = {Martyna, Glenn J. and Tobias, Douglas J. and Klein, Michael L.},
	month = sep,
	year = {1994},
	pages = {4177--4189},
}

@article{vandevondele_gaussian_2007,
	title = {Gaussian basis sets for accurate calculations on molecular systems in gas and condensed phases},
	volume = {127},
	issn = {0021-9606, 1089-7690},
	url = {http://aip.scitation.org/doi/10.1063/1.2770708},
	doi = {10.1063/1.2770708},
	language = {en},
	number = {11},
	urldate = {2022-09-23},
	journal = {The Journal of Chemical Physics},
	author = {VandeVondele, Joost and Hutter, Jürg},
	month = sep,
	year = {2007},
	pages = {114105},
}

@article{kuhne_cp2k_2020,
	title = {{CP2K}: {An} electronic structure and molecular dynamics software package - {Quickstep}: {Efficient} and accurate electronic structure calculations},
	volume = {152},
	issn = {0021-9606},
	url = {https://doi.org/10.1063/5.0007045},
	doi = {10.1063/5.0007045},
	number = {19},
	journal = {The Journal of Chemical Physics},
	author = {Kühne, Thomas D. and Iannuzzi, Marcella and Del Ben, Mauro and Rybkin, Vladimir V. and Seewald, Patrick and Stein, Frederick and Laino, Teodoro and Khaliullin, Rustam Z. and Schütt, Ole and Schiffmann, Florian and Golze, Dorothea and Wilhelm, Jan and Chulkov, Sergey and Bani-Hashemian, Mohammad Hossein and Weber, Valéry and Borštnik, Urban and Taillefumier, Mathieu and Jakobovits, Alice Shoshana and Lazzaro, Alfio and Pabst, Hans and Müller, Tiziano and Schade, Robert and Guidon, Manuel and Andermatt, Samuel and Holmberg, Nico and Schenter, Gregory K. and Hehn, Anna and Bussy, Augustin and Belleflamme, Fabian and Tabacchi, Gloria and Glöß, Andreas and Lass, Michael and Bethune, Iain and Mundy, Christopher J. and Plessl, Christian and Watkins, Matt and VandeVondele, Joost and Krack, Matthias and Hutter, Jürg},
	month = may,
	year = {2020},
	pages = {194103},
}

@article{van_setten_pseudodojo_2018,
	title = {The {PseudoDojo}: Training and grading a 85 element optimized norm-conserving pseudopotential table},
	volume = {226},
	issn = {0010-4655},
	url = {https://www.sciencedirect.com/science/article/pii/S0010465518300250},
	doi = {10.1016/j.cpc.2018.01.012},
	shorttitle = {The {PseudoDojo}},
	pages = {39--54},
	journal = {Computer Physics Communications},
	author = {van Setten, M. J. and Giantomassi, M. and Bousquet, E. and Verstraete, M. J. and Hamann, D. R. and Gonze, X. and Rignanese, G. -M.},
	year = {2018},
}

@article{monkhorst_special_1976,
	title = {Special points for Brillouin-zone integrations},
	volume = {13},
	url = {https://link.aps.org/doi/10.1103/PhysRevB.13.5188},
	doi = {10.1103/PhysRevB.13.5188},
	pages = {5188--5192},
	number = {12},
	journal = {Physical Review B},
	shortjournal = {Phys. Rev. B},
	author = {Monkhorst, Hendrik J. and Pack, James D.},
	urldate = {2022-05-23},
	year = {1976},
}

@article{supka_aflow_2017,
	title = {{AFLOW$\pi$}: A minimalist approach to high-throughput \textit{ab initio} calculations including the generation of tight-binding hamiltonians},
	volume = {136},
	issn = {0927-0256},
	url = {https://www.sciencedirect.com/science/article/pii/S0927025617301970},
	doi = {10.1016/j.commatsci.2017.03.055},
	shorttitle = {{AFLOWπ}},
	pages = {76--84},
	journal = {Computational Materials Science},
	author = {Supka, Andrew R. and Lyons, Troy E. and Liyanage, Laalitha and D’Amico, Pino and Al Rahal Al Orabi, Rabih and Mahatara, Sharad and Gopal, Priya and Toher, Cormac and Ceresoli, Davide and Calzolari, Arrigo and Curtarolo, Stefano and Nardelli, Marco Buongiorno and Fornari, Marco},
	year = {2017},
}

@article{giannozzi_quantum_2009,
	title = {{QUANTUM} {ESPRESSO}: a modular and open-source software project for quantum simulations of materials},
	volume = {21},
	issn = {0953-8984},
	url = {https://doi.org/10.1088/0953-8984/21/39/395502},
	doi = {10.1088/0953-8984/21/39/395502},
	shorttitle = {{QUANTUM} {ESPRESSO}},
	pages = {395502},
	number = {39},
	journal = {Journal of Physics: Condensed Matter},
	shortjournal = {J. Phys.: Condens. Matter},
	author = {Giannozzi, Paolo and Baroni, Stefano and Bonini, Nicola and Calandra, Matteo and Car, Roberto and Cavazzoni, Carlo and Ceresoli, Davide and Chiarotti, Guido L. and Cococcioni, Matteo and Dabo, Ismaila and Corso, Andrea Dal and Gironcoli, Stefano de and Fabris, Stefano and Fratesi, Guido and Gebauer, Ralph and Gerstmann, Uwe and Gougoussis, Christos and Kokalj, Anton and Lazzeri, Michele and Martin-Samos, Layla and Marzari, Nicola and Mauri, Francesco and Mazzarello, Riccardo and Paolini, Stefano and Pasquarello, Alfredo and Paulatto, Lorenzo and Sbraccia, Carlo and Scandolo, Sandro and Sclauzero, Gabriele and Seitsonen, Ari P. and Smogunov, Alexander and Umari, Paolo and Wentzcovitch, Renata M.},
	urldate = {2022-05-20},
	year = {2009},
	langid = {english},
}

@article{carnevali_nanoscale_2025,
	title = {Nanoscale size effects in $\alpha$-{FAPbI}$_3$ evinced by large-scale ab initio simulations},
	volume = {16},
	rights = {2025 The Author(s)},
	issn = {2041-1723},
	url = {https://www.nature.com/articles/s41467-025-61351-7},
	doi = {10.1038/s41467-025-61351-7},
	pages = {6825},
	number = {1},
	journal = {Nature Communications},
	shortjournal = {Nat Commun},
	author = {Carnevali, Virginia and Agosta, Lorenzo and Slama, Vladislav and Lempesis, Nikolaos and Vezzosi, Andrea and Rothlisberger, Ursula},
	urldate = {2025-07-24},
	year = {2025},
	langid = {english},
}

@phdthesis{gardner_ferroelectricity,
  author       = {Gardner, Jonathan},
  title        = {Ferroelectricity in empty tetragonal tungsten bronzes},
  school       = {University of St Andrews},
  year         = {2017},
  type         = {PhD thesis},
}

@article{roshchupkin_study_2009,
	title = {Study of {LiNbO3} and {LiTaO3} ferroelectric domain structures using high-resolution x-ray diffraction under application of external electric field},
	volume = {105},
	issn = {0021-8979},
	url = {https://doi.org/10.1063/1.3066481},
	doi = {10.1063/1.3066481},
	number = {2},
	urldate = {2025-03-20},
	journal = {Journal of Applied Physics},
	author = {Roshchupkin, D. V. and Irzhak, D. V. and Antipov, V. V.},
	month = jan,
	year = {2009},
	pages = {024112},
}

@article{olsen_role_2017,
	title = {Role of {Lone} {Pair} {Cations} in {Ferroelectric} {Tungsten} {Bronzes}},
	volume = {29},
	issn = {0897-4756},
	url = {https://doi.org/10.1021/acs.chemmater.7b01817},
	doi = {10.1021/acs.chemmater.7b01817},
	number = {15},
	urldate = {2023-01-09},
	journal = {Chemistry of Materials},
	author = {Olsen, Gerhard Henning and Sørby, Magnus Helgerud and Selbach, Sverre Magnus and Grande, Tor},
	month = aug,
	year = {2017},
	note = {Publisher: American Chemical Society},
	pages = {6414--6424},
}

@article{KASPERA201776,
title = {Alkali tungsten bronzes as soot oxidation catalysts: The key role of electrodonor properties of catalytic surface},
journal = {Catalysis Communications},
volume = {98},
pages = {76-80},
year = {2017},
issn = {1566-7367},
doi = {https://doi.org/10.1016/j.catcom.2017.05.009},
url = {https://www.sciencedirect.com/science/article/pii/S1566736717302030},
author = {Wojciech Kaspera and Sebastian Zieliński and Andrzej Kotarba}
}

@article{BIERSTEDT196625,
title = {Superconductivity of some new hexagonal tungsten bronzes},
journal = {Solid State Communications},
volume = {4},
number = {1},
pages = {25-26},
year = {1966},
issn = {0038-1098},
doi = {https://doi.org/10.1016/0038-1098(66)90097-4},
url = {https://www.sciencedirect.com/science/article/pii/0038109866900974},
author = {P.E. Bierstedt and T.A. Bither and F.J. Darnell}
}

@article{tegg_review_2023,
	title = {A {Review} of {Alkali} {Tungsten} {Bronze} {Nanoparticles} for {Applications} in {Plasmonics}},
	volume = {18},
	issn = {1557-1963},
	url = {https://doi.org/10.1007/s11468-022-01749-x},
	doi = {10.1007/s11468-022-01749-x},
	language = {en},
	number = {1},
	urldate = {2024-09-30},
	journal = {Plasmonics},
	author = {Tegg, Levi and Keast, Vicki J.},
	month = feb,
	year = {2023},
	pages = {49--71},
}

@Article{D0RA06028C,
author ="Wang, Sanjun and Xiao, Wen-bo and Wang, Fei",
title  ="Structural{,} electronic{,} and optical properties of cubic formamidinium lead iodide perovskite: a first-principles investigation",
journal  ="RSC Adv.",
year  ="2020",
volume  ="10",
issue  ="54",
pages  ="32364-32369",
publisher  ="The Royal Society of Chemistry",
doi  ="10.1039/D0RA06028C",
url  ="http://dx.doi.org/10.1039/D0RA06028C"}

@article{Pero,
  author = {Schmidt-Mende,Lukas  and Dyakonov,Vladimir  and Olthof,Selina  and Ünlü,Feray  and Lê,Khan Moritz Trong  and Mathur,Sanjay  and Karabanov,Andrei D.  and Lupascu,Doru C.  and Herz,Laura M.  and Hinderhofer,Alexander  and Schreiber,Frank  and Chernikov,Alexey  and Egger,David A.  and Shargaieva,Oleksandra  and Cocchi,Caterina  and Unger,Eva  and Saliba,Michael  and Byranvand,Mahdi Malekshahi  and Kroll,Martin  and Nehm,Frederik  and Leo,Karl  and Redinger,Alex  and Höcker,Julian  and  Kirchartz,Thomas  and Warby,Jonathan  and Gutierrez-Partida,Emilio  and Neher,Dieter  and Stolterfoht,Martin  and Würfel,Uli  and Unmüssig,Moritz  and Herterich,Jan  and Baretzky,Clemens  and Mohanraj,John  and Thelakkat,Mukundan  and Maheu,Clément  and Jaegermann,Wolfram  and Mayer,Thomas  and Rieger,Janek  and Fauster,Thomas  and Niesner,Daniel  and Yang,Fengjiu  and Albrecht,Steve  and Riedl,Thomas  and Fakharuddin,Azhar  and Vasilopoulou,Maria  and Vaynzof,Yana  and Moia,Davide  and   Maier,Joachim  and Franckevičius,Marius  and Gulbinas,Vidmantas  and Kerner,Ross A.  and Zhao,Lianfeng  and Rand,Barry P.  and Glück,Nadja  and Bein,Thomas  and Matteocci,Fabio  and Castriotta,Luigi Angelo  and Di Carlo,Aldo  and Scheffler,Matthias  and Draxl,Claudia },
  title = {Roadmap on organic–inorganic hybrid perovskite semiconductors and devices},
  journal = {APL Materials},
  volume = {9},
  number = {10},
  pages = {109202},
  year = {2021},
  doi = {10.1063/5.0047616},
}

@article{Olsen,
  title = {Origin of ferroelectric polarization in tetragonal tungsten-bronze-type oxides},
  author = {Olsen, Gerhard Henning and Aschauer, Ulrich and Spaldin, Nicola A. and Selbach, Sverre Magnus and Grande, Tor},
  journal = {Phys. Rev. B},
  volume = {93},
  issue = {18},
  pages = {180101},
  numpages = {5},
  year = {2016},
  doi = {10.1103/PhysRevB.93.180101},
}

@article{Shannon,
  author = {JR. D. Shannon},
  title = {Revised effective ionic radii and systematic studies of inter-atomic distances in halides and chalcogenidess},
  journal = {Acta Crystallogr.},
  volume = {32},
  pages = {751-767},
  year = {1976},
  doi = {10.1107/S0567739476001551},
}

@article{Halide,
  author = {Jena, Ajay Kumar and Kulkarni, Ashish and Miyasaka, Tsutomu},
  title = {Halide Perovskite Photovoltaics: Background, Status, and Future Prospects},
  journal = {Chemical Reviews},
  volume = {119},
  number = {5},
  pages = {3036-3103},
  year = {2019},
  doi = {10.1021/acs.chemrev.8b00539},
}

@article{PeroRev,
  author = {Schmidt-Mende,Lukas  and Dyakonov,Vladimir  and Olthof,Selina  and Ünlü,Feray  and Lê,Khan Moritz Trong  and Mathur,Sanjay  and Karabanov,Andrei D.  and Lupascu,Doru C.  and Herz,Laura M.  and Hinderhofer,Alexander  and Schreiber,Frank  and Chernikov,Alexey  and Egger,David A.  and Shargaieva,Oleksandra  and Cocchi,Caterina  and Unger,Eva  and Saliba,Michael  and Byranvand,Mahdi Malekshahi  and Kroll,Martin  and Nehm,Frederik  and Leo,Karl  and Redinger,Alex  and Höcker,Julian  and  Kirchartz,Thomas  and Warby,Jonathan  and Gutierrez-Partida,Emilio  and Neher,Dieter  and Stolterfoht,Martin  and Würfel,Uli  and Unmüssig,Moritz  and Herterich,Jan  and Baretzky,Clemens  and Mohanraj,John  and Thelakkat,Mukundan  and Maheu,Clément  and Jaegermann,Wolfram  and Mayer,Thomas  and Rieger,Janek  and Fauster,Thomas  and Niesner,Daniel  and Yang,Fengjiu  and Albrecht,Steve  and Riedl,Thomas  and Fakharuddin,Azhar  and Vasilopoulou,Maria  and Vaynzof,Yana  and Moia,Davide  and   Maier,Joachim  and Franckevičius,Marius  and Gulbinas,Vidmantas  and Kerner,Ross A.  and Zhao,Lianfeng  and Rand,Barry P.  and Glück,Nadja  and Bein,Thomas  and Matteocci,Fabio  and Castriotta,Luigi Angelo  and Di Carlo,Aldo  and Scheffler,Matthias  and Draxl,Claudia },
  title = {Roadmap on organic–inorganic hybrid perovskite semiconductors and devices},
  journal = {APL Materials},
  volume = {9},
  number = {10},
  pages = {109202},
  year = {2021},
  doi = {10.1063/5.0047616},
}

@article{MF,
  title = {Interplay between $A$-site and $B$-site driven instabilities in perovskites},
  author = {Ghita, M. and Fornari, M. and Singh, D. J. and Halilov, S. V.},
  journal = {Phys. Rev. B},
  volume = {72},
  issue = {5},
  pages = {054114},
  numpages = {5},
  year = {2005},
  doi = {10.1103/PhysRevB.72.054114},
}

@article{toll,
	doi = {10.1143/jjap.40.5999},
	year = 2001,
	volume = {40},
	number = {Part 1, No. 10},
	pages = {5999--6002},
  journal = {Ceramics},
	author = {Richard E. Eitel and Clive A. Randall and Thomas R. Shrout and Paul W. Rehrig and Wes Hackenberger and Seung-Eek Park},
	title = {New High Temperature Morphotropic Phase Boundary Piezoelectrics Based on Bi(Me)O$_3${\textendash}{PbTiO}$_3$},
}

@article{Gold,
  author = {V.M. Goldschmidt},
  title = {Die Gesetze der Krystallochemie},
  journal = {Naturwissenschaften},
  volume = {14},
  pages = {477-485},
  year = {1926},
  doi = {10.1007/BF01507527},
}

@article{PT1,
  title = {Finite-Temperature Properties of $\mathrm{Pb}({{\mathrm{Zr}}_{1\ensuremath{-}\mathit{x}}\mathrm{Ti}}_{\mathit{x}}){O}_{3}$ Alloys from First Principles},
  author = {Bellaiche, L. and Garc\'{\i}a, Alberto and Vanderbilt, David},
  journal = {Phys. Rev. Lett.},
  volume = {84},
  issue = {23},
  pages = {5427--5430},
  numpages = {0},
  year = {2000},
  month = {Jun},
  publisher = {American Physical Society},
  doi = {10.1103/PhysRevLett.84.5427},
  url = {https://link.aps.org/doi/10.1103/PhysRevLett.84.5427}
}

@article{PT,
  title = {First-principles investigation of ferroelectricity in perovskite compounds},
  author = {King-Smith, R. D. and Vanderbilt, David},
  journal = {Phys. Rev. B},
  volume = {49},
  issue = {9},
  pages = {5828--5844},
  numpages = {0},
  year = {1994},
  month = {Mar},
  publisher = {American Physical Society},
  doi = {10.1103/PhysRevB.49.5828},
  url = {https://link.aps.org/doi/10.1103/PhysRevB.49.5828}
}

@article{Guo1,
author = {Guo,R.  and Bhalla,A. S.  and Randall,C. A.  and Cross,L. E. },
title = {Dielectric and pyroelectric properties of the morphotropic phase boundary lead barium niobate (PBN) single crystals at low temperature (10–300 K)},
journal = {Journal of Applied Physics},
volume = {67},
number = {10},
pages = {6405-6410},
year = {1990},
doi = {10.1063/1.345164},
}

@ARTICLE{Labbe,
  author = {P. Labb{\'e} and M. Frey and G. Allias},
  title = {Nouvelles donn{\'e}es structurales sur la vari{\'e}te ferro{\'e}lectrique du m{\'e}taniobate de plomb PbNb$_2$O$_6$},
  journal = {Acta {C}rystallogr. {B}},
  year = {1973},
  pages = {12204–2210},
  number = {26},
  doi = {10.1021/ja0276321},
}

@BOOK{Lines,
  author = {M. E. Lines and A. M. Glass},
  title = {Principles and applications of ferroelectrics and related materials},
  publisher = {{O}xford: {O}xford {U}niversity {P}ress},
  year = {2001},
  doi = {10.1093/acprof:oso/9780198507789.001.0001}
}

@BOOK{MPB,
  author = {Ruyan Guo and K. M. Nair and Winnie Wong-Ng and Amar S. Bhalla and Dwight Vieland and D. Suvorov and Carl Wu and S.-I. Hirano},
  title = {Morphotropic {P}hase {B}oundary {P}erovskites, {H}igh {S}train {P}iezoelectrics, and {D}ielectric {C}eramics},
  publisher = {Wiley},
  year = {2003},
  volume = {100},
  pages = {535}
}

@BOOK{Hyde,
  author = {B. G. Hyde and S. Andersson},
  title = {Inorganic crystal structures},
  publisher = {Wiley},
  year = {1990},
  doi = {10.1002/crat.2170250611}
}

@article{Randall,
  title={Microstructure-property relations in tungsten bronze lead barium niobate,{P}b{$_{1-x}$}{B}a{$_x$}{N}b{$_2$}{O}{$_6$}},
  volume={6},
  DOI={10.1557/JMR.1991.1720},
  number={8},
  journal={Journal of Materials Research},
  publisher={Cambridge University Press},
  author={Randall, C.A. and Guo, R. and Bhalla, A.S. and Cross, L.E.},
  year={1991},
  pages={1720–1728}
}

@article{Guo,
author = {Guo,R.  and Bhalla,A. S.  and Randall,C. A.  and Chang,Z. P.  and Cross,L. E. },
title = {Polarization mechanisms of morphotropic phase boundary lead barium niobate (PBN) compositions},
journal = {Journal of Applied Physics},
volume = {67},
number = {3},
pages = {1453-1460},
year = {1990},
doi = {10.1063/1.345651},
}

@article{Shrout,
author = { T. R.   Shrout  and  L. E.   Cross  and  D. A.   Hukin },
title = {Ferroelectric properties of tungsten bronze lead barium niobate (PBN) single crystals},
journal = {Ferroelectrics},
volume = {44},
number = {1},
pages = {325-330},
year  = {1982},
publisher = {Taylor & Francis},
doi = {10.1080/00150198308260680}
}

@article{Shrout1,
author = { T. R.   Shrout  and  Huanchu   Chen  and  L. E.   Cross },
title = {Dielectric and piezoelectric properties of {P}b{$_{1-x}$}{B}a{$_x$}{N}b{$_2$}{O}{$_6$} ferroelectric tungsten bronze crystals},
journal = {Ferroelectrics},
volume = {74},
number = {1},
pages = {317-324},
year  = {1987},
publisher = {Taylor & Francis},
doi = {10.1080/00150198708201313},
}

@article{Lane,
author = { R. Lane  and  D. L. Mack and K. R. Brown},
journal = {Trans. J. Brit. Ceramic Soc.},
volume = {71},
number = {11},
year  = {1972},
}

@article{GOODMAN,
author = {Goodman, G.},
title = {Ferroelectric Properties of Lead Metaniobate},
journal = {Journal of the American Ceramic Society},
volume = {36},
number = {11},
pages = {368-372},
doi = {10.1111/j.1151-2916.1953.tb12820.x},
year = {1953}
}

@article{Francombe,
author = { M. H. Francombe},
title = {Polymorphism in lead metaniobate},
journal = {Acta Crystallogr.},
volume = {9},
pages = {683-684},
doi = {10.1107/S0365110X56001856},
year = {1956}
}

@article{Francombe2,
author = { M. H. Francombe},
title = {The relation between structure and ferroelectricity in lead barium and barium strontium niobates},
journal = {Acta Crystallogr.},
volume = {13},
pages = {131-140},
doi = {10.1107/S0365110X60000285},
year = {1960}
}

@article{Sahini,
author = { M.G. Sahini and Grande, T. and Fraygola, B. and Biancoli, A. and Damjanovic, D. and Setter,N.},
title = {Solid Solutions of Lead Metaniobate-Stabilization of the Ferroelectric Polymorph and the Effect on the Lattice Parameters, Dielectric, Ferroelectric, and Piezoelectric Propertie},
journal = {Journal of the American Ceramic Society},
volume = {97},
issue = {1},
pages = {220-227},
doi = {10.1111/jace.12628},
year = {2014}
}

@article{Cross,
author = { L. Eric   Cross },
title = {Relaxor ferroelectrics},
journal = {Ferroelectrics},
volume = {76},
number = {1},
pages = {241-267},
year  = {1987},
publisher = {Taylor & Francis},
doi = {10.1080/00150198708016945},
}

\end{document}


\begin{figure*}[h!]
  \centering
  \includegraphics[scale=0.35]{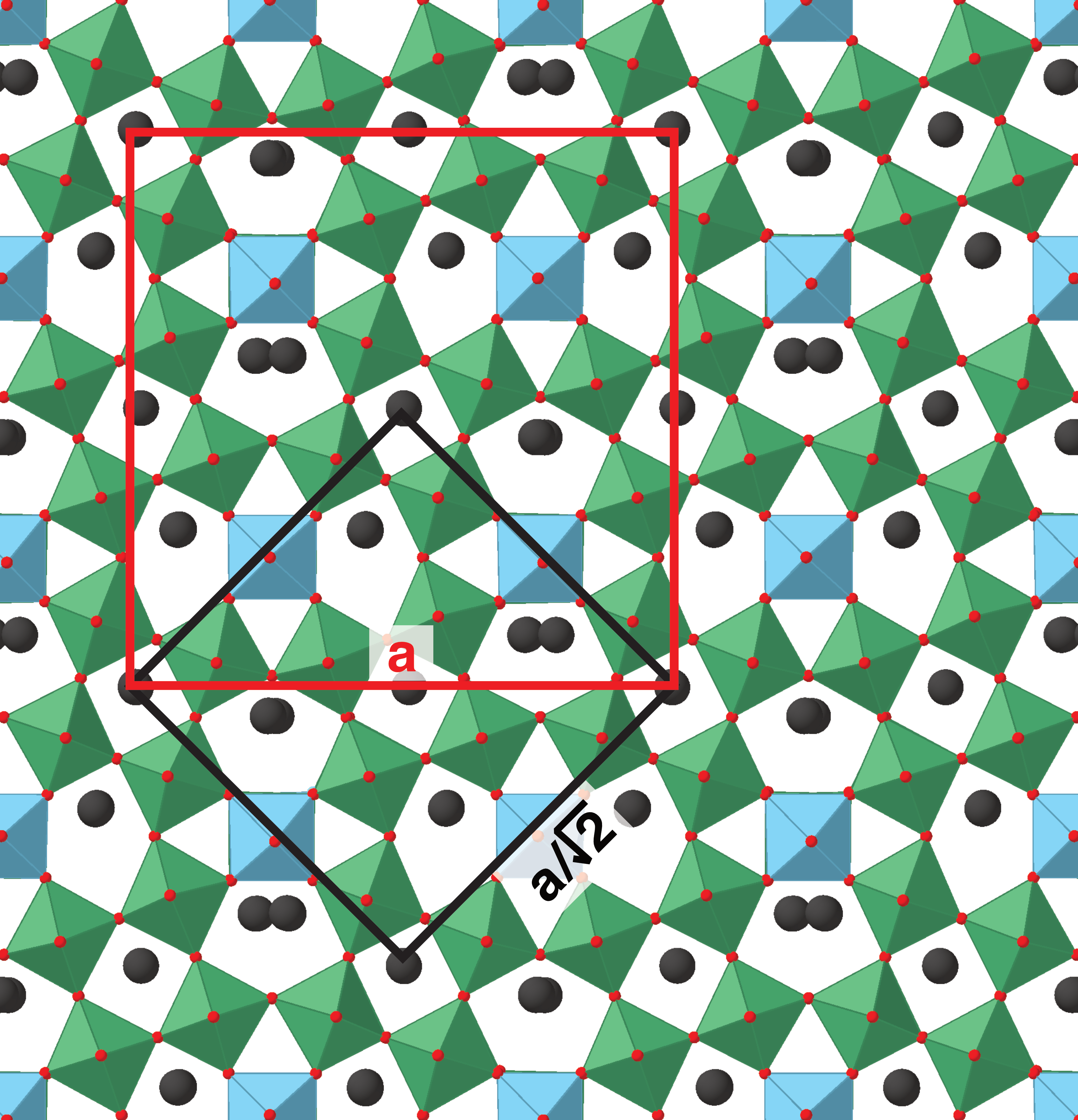}
  \caption{TTB simulation cells in relation to the orthorhombic symmetry (cell-1, red solid line) and the tetragonal symmetry (cell-2, black solid line). The lattice vectors are expressed in relation to the one of cell-1 (a). Pb, O, and Nb are represented by gray balls, red balls, and green/light blue octahedra, respectively.}
  \label{fig:S1}
\end{figure*}

\newpage

\begin{figure*}[h!]
  \centering
  \includegraphics[scale=0.7]{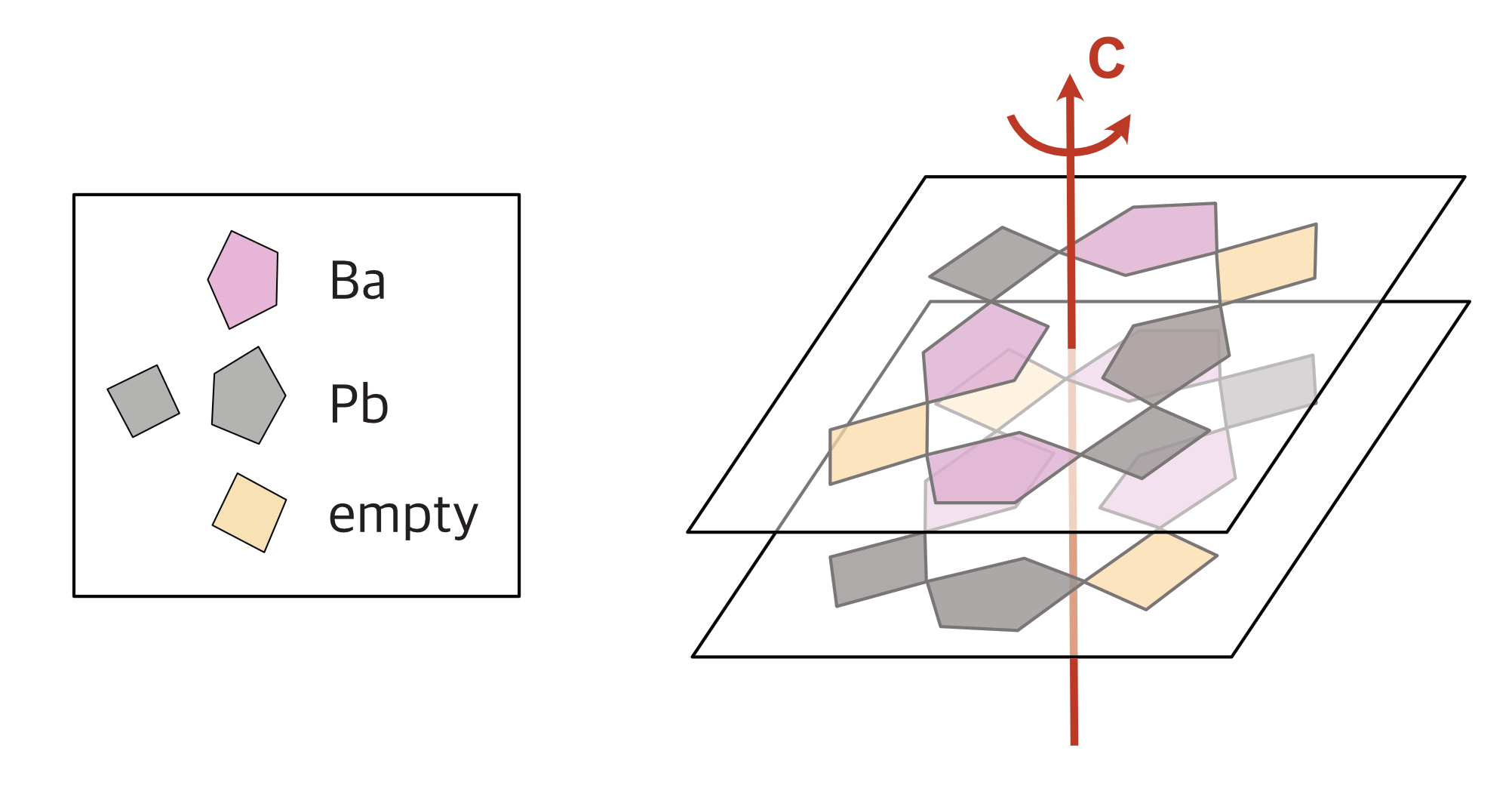}
  \caption{Structural model for Pb$_{1-x}$Ba$_x$Nb$_2$O$_6$. The models are built considering the double cell periodicity along the $c$ axis, the in-plane checkerboard pattern, and the screw axis along the $c$ direction for the $A^{XII}$ vacancy sites.}
  \label{fig:S2}
\end{figure*}

\newpage

\begin{figure*}[h!]
  \centering
  \includegraphics[scale=0.4]{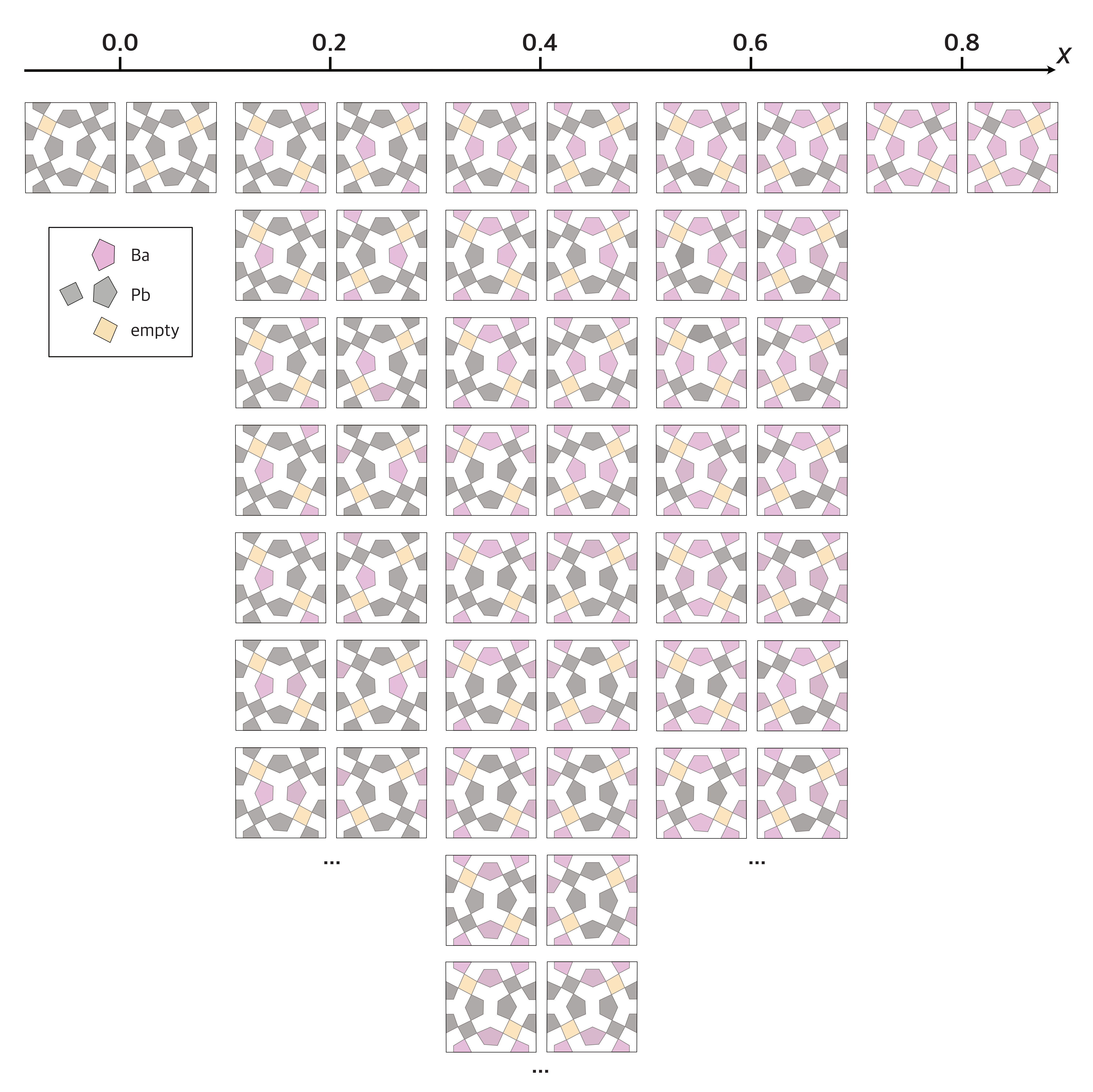}
  \caption{Structural models for Pb$_{1-x}$Ba$_x$Nb$_2$O$_6$. For each $x$, a sample of models is stacked in column; the first and the second layer of the model are represented side by side. The models are sketched according to the filling of the A$^{XII}$ and A$^{XV}$ sites. On top, the formation energy of all the computed models at each $x$.}
  \label{fig:S3}
\end{figure*}

\newpage

\begin{figure*}[h!]
  \centering
  \includegraphics[scale=0.55]{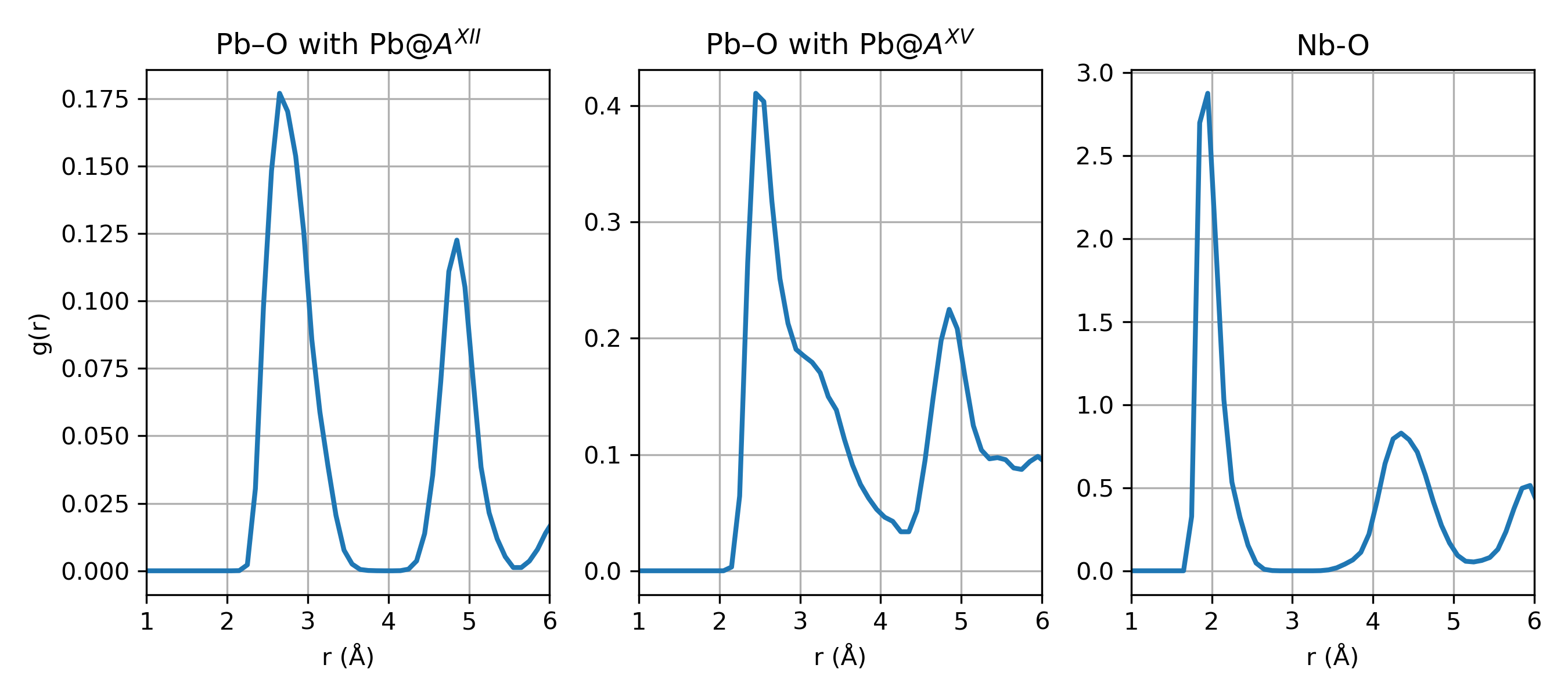}
  \caption{Radial distribution functions of the different elements for the orthorhombic PbNb$_2$O$_6$ TTB. AIMD was run for 5 ps in the NVE ensemble at 300 K, with the Nb statistic taken every 25 fs.}
  \label{fig:S4}
\end{figure*}

\newpage

\newpage

\begin{figure*}[h!]
  \centering
  \includegraphics[scale=0.55]{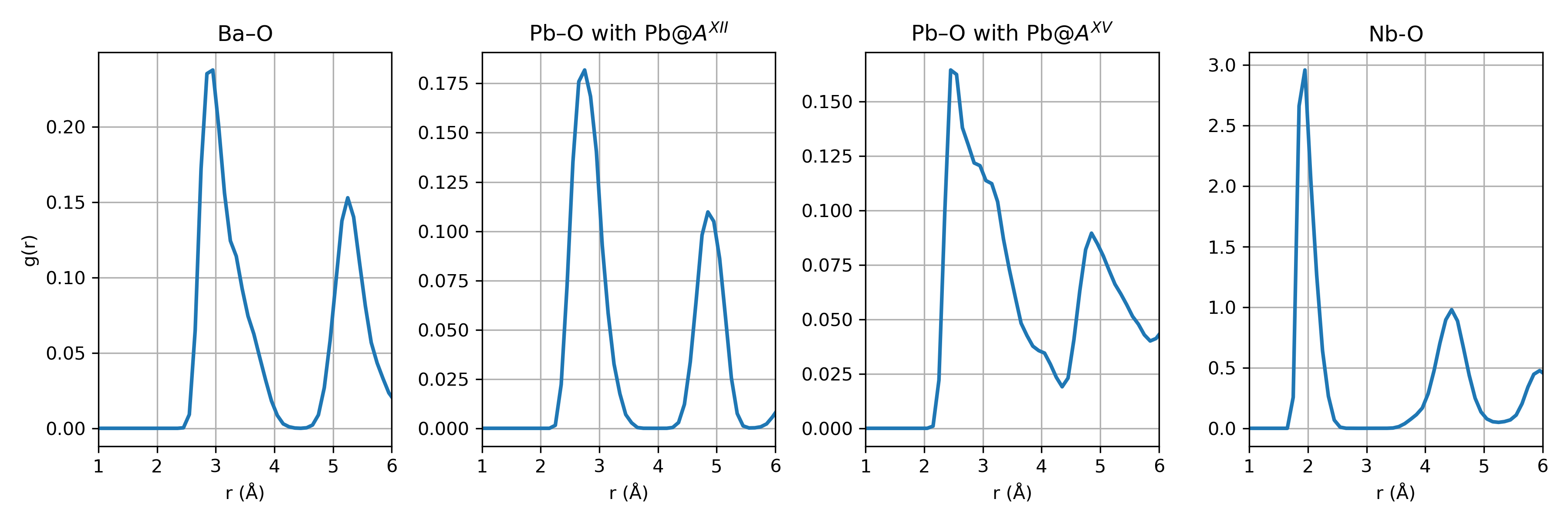}
  \caption{Radial distribution functions of the different elements for the tetragonal Pb$_{0.23}$Ba$_{0.67}$Nb$_2$O$_6$ TTB. AIMD was run for 5 ps in the NVE ensemble at 300 K, with the Nb statistic taken every 25 fs.}
  \label{fig:S5}
\end{figure*}

\newpage

\begin{figure*}[h!]
  \centering
  \includegraphics[scale=0.64]{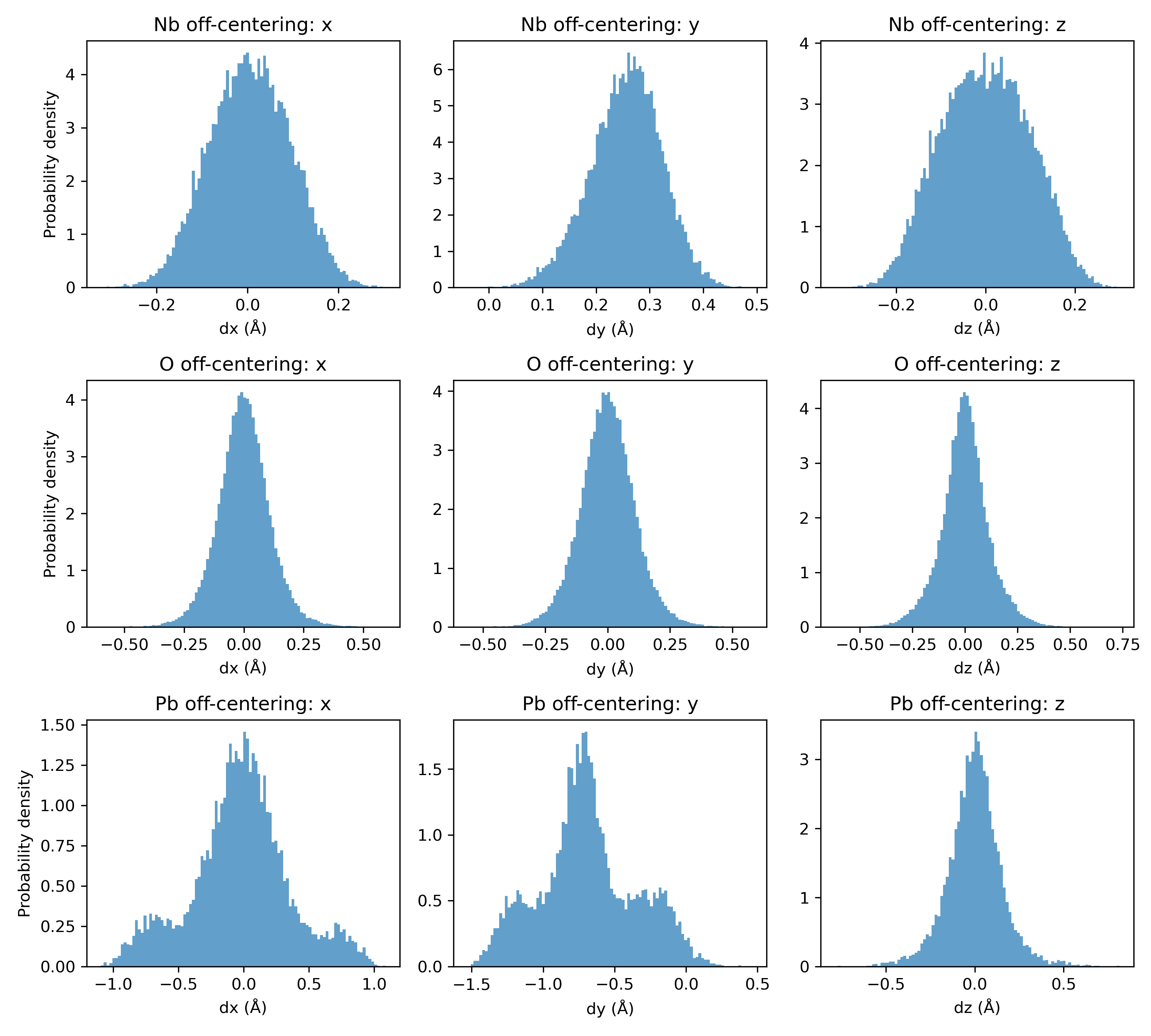}
  \caption{Element off-centering distributions along x-, y-, and z- directions for the orthorhombic PbNb$_2$O$_6$ TTB. AIMD was run for 5 ps in the NVE ensemble at 300 K, with the Nb statistic taken every 25 fs.}
  \label{fig:S6}
\end{figure*}

\newpage

\begin{figure*}[h!]
  \centering
  \includegraphics[scale=0.64]{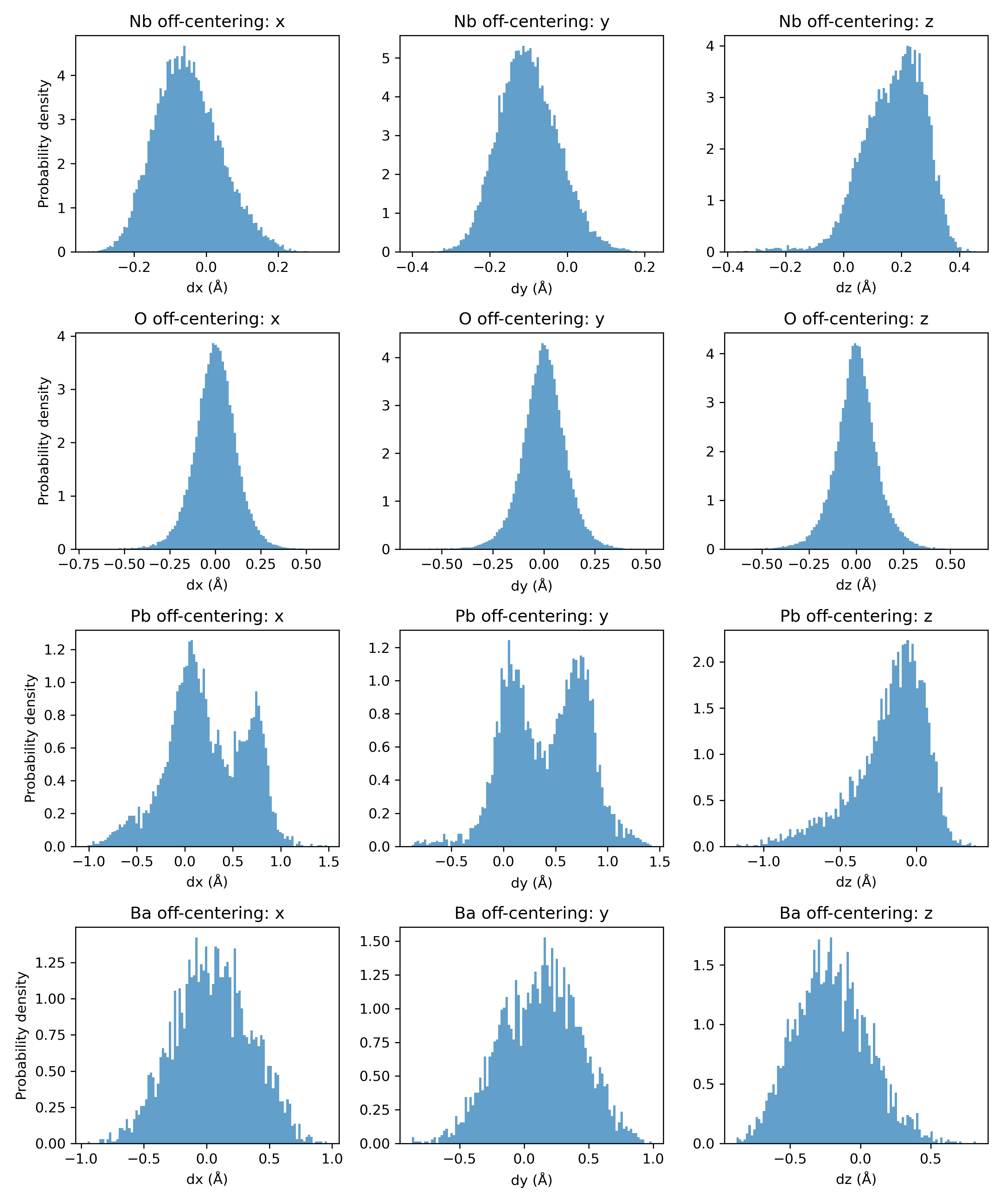}
  \caption{Element off-centering distributions along x-, y-, and z- directions for the tetragonal Pb$_{0.23}$Ba$_{0.67}$Nb$_2$O$_6$ TTB. AIMD was run for 5 ps in the NVE ensemble at 300 K, with the Nb statistic taken every 25 fs.}
  \label{fig:S5}
\end{figure*}

\newpage

\begin{table*}[h!]
    \centering
    \begin{tabular}{c|ccccc}
      \textbf{x}              & 0 &0.2 & 0.4 & 0.6 & 0.8 \\
      \textbf{\# conf}        & 1        & 196                 & 1225                & 196                 & 1      \\
      \textbf{\# conf comput} & 1        & 5                   & 10                  & 5                   & 1      \\ 
      $\bm{E_f /atom}$ \textbf{(eV)}    & -0.064   & -0.096 $\pm$ 0.001  & -0.125 $\pm$ 0.001  & -0.158 $\pm$ 0.001  & -0.189 \\ 
    \end{tabular}
    \caption{Total number of configurations (\# conf), simulated configurations (\# conf comput), and formation energy at given Ba concentration (x) averaged on the computed configurations. }
    \label{TabS1:conf}
\end{table*}

\newpage

\begin{table*}[h!]
    \centering
    \resizebox{\textwidth}{!}{\begin{tabular}{c|ccccc}
                              & \textbf{element} & \textbf{dx (\AA)} & \textbf{dy (\AA)} & \textbf{dz (\AA)}  & \bm$|$\textbf{d}\bm$|$ \textbf{(\AA)}  \\
      \hline
      \multirow{2}{*}{PbNi$_5$O$_6$}                     & Nb & $0.006 \pm 0.089$ & $0.256 \pm 0.067$ & $0.0003 \pm 0.0985$ & $0.289 \pm 0.061$ \\
                                                         & Pb & $-0.027 \pm 0.378$ & $-0.688 \pm 0.338$ & $0.0002 \pm 0.1565$ & $0.808 \pm 0.322$ \\
      \hline
      \multirow{3}{*}{Pb$_{0.23}$Ba$_{0.67}$Ni$_5$O$_6$} & Nb & $-0.050 \pm 0.090$ & $-0.099 \pm 0.076$ & $0.174 \pm 0.106$ & $0.251 \pm 0.069$ \\
                                                         & Pb & $0.228 \pm 0.407$  & $0.406 \pm 0.378$  & $-0.172 \pm 0.235$ & $0.709 \pm 0.328$ \\
                                                         & Ba & $0.046 \pm 0.307$  & $0.126 \pm 0.306$  & $-0.203 \pm 0.262$ & $0.532 \pm 0.180$ \\
    \end{tabular}}
    \caption{Nb, Pb, and Ba off-centering averages along x-, y-, and z- directions for the orthorhombic PbNb$_2$O$_6$ and the tetragonal Pb$_{0.23}$Ba$_{0.67}$Nb$_2$O$_6$ TTB.  AIMD was run for 5 ps in the NVE ensemble at 300 K, with the Nb statistic taken every 25 fs. }
    \label{TabS2:disp}
\end{table*}